\documentclass[preprint,12pt,fleqn]{elsarticle}
\usepackage{amssymb}
\usepackage{amsfonts}
\usepackage[english]{babel}
\usepackage{euscript}
\usepackage{bbm}
\usepackage{xfrac}

\usepackage{color}
\usepackage{mathrsfs}
\usepackage{calrsfs}
\usepackage{amsbsy}

\newcommand{\eqref}[1]{(\ref{#1})}

\newcommand{\E}{\mathrm{E}}
\newcommand{\J}{\mathrm{J}}

\newcounter{example_counter}

\newcounter{definition_counter}
\newcounter{theorem_counter}
\newcounter{remark_counter}
\newcommand{\uu}{\pmb{u}}
\newcommand{\qq}{\pmb{q}}

\newcommand{\ee}{\pmb{E}}
\newcommand{\vv}{\pmb{v}}
\newcommand{\ww}{\pmb{w}}

\newcommand{\crl}{\mathrm{curl}\,}
\newcommand{\ccrl}{\mathrm{curl^2}\,}

\begin{document}

\begin{frontmatter}

\title{Lax representations  for the  Navier--Stokes  equations}

\author{Oleg I. Morozov}
\ead{oimorozov@gmail.com}

\begin{abstract}
We construct a family of Lax representations  for the three-dimensional incompressible Navier--Stokes equations in vorticity form. The key idea is to introduce an auxiliary vector field, termed the \emph{vorticity-diffusion potential}, which recasts the viscous Laplacian diffusion term as a Lie-bracket advection term. On solutions of the defining equation for this potential, the viscous vorticity equation takes the form of a pure Lie-transport equation along a modified divergence-free velocity field, while the defining constraint itself is encoded as the curl-free compatibility condition for an auxiliary scalar potential. We further show that the reduction of this three-dimensional system to two spatial dimensions yields a Lax representation for the two-dimensional Navier--Stokes vorticity equation, which we generalize to a family parametrized by an arbitrary smooth function of one variable.
\end{abstract}

\begin{keyword}
Navier--Stokes equations \sep vorticity form \sep Lax representation \sep vorticity-diffusion potential

\MSC[2020] 35Q30 \sep 37K10 \sep 58J70
\end{keyword}

\end{frontmatter}


\section{Introduction}

The three-dimensional Navier--Stokes equations are fundamental to the mathematical description of viscous incompressible fluid flows \cite{LandauLifshits,ChorinMarsden}. They arise across a wide range of problems in fluid mechanics and continue to attract intense interest from the viewpoints of mathematical analysis, differential geometry, and mathematical physics. Despite their central role and the extensive body of research devoted to them, the analytical and geometric structure of these equations remains far from completely understood. One approach to investigating their structural properties is to seek reformulations that uncover hidden algebraic or geometric features, which in turn may provide tools for constructing exact or particular solutions, symmetries, conservation laws, and related analytical objects.

In the theory of nonlinear partial differential equations, \emph{Lax representations} provide a powerful geometric framework of this kind \cite{AblowitzClarkson1991,Konopelchenko1987,NovikovManakovPitaevskiyZakharov1984,Zakharov82}. A Lax representation encodes a given nonlinear system as the compatibility (integrability) condition for an auxiliary overdetermined system of differential equations involving additional dependent variables, often called \emph{potentials} or \emph{pseudopotentials} \cite{WE}. Within the geometric theory of differential equations, such auxiliary systems are formalized as \emph{differential coverings}   \cite{KrasilshchikVinogradov1984,KrasilshchikVinogradov1989,VK1999}  and serve as a starting point for constructing B\"{a}cklund and Darboux transformations \cite{RogersShadwick1982,MatveevSalle1991,BacklundDarboux2001}, nonlocal symmetries, and hierarchies of conservation laws.  Finding a Lax representation can thus reveal structural properties of a PDE system that are not immediately apparent from its original formulation.

For the inviscid vorticity equations,  the nonlinear advection term  is purely a Lie bracket of divergence-free vector fields (or a Jacobian in two dimensions), and Lax representations can be constructed using the Lie algebra of volume-preserving vector fields and its outer derivations; see \cite{Li2001,LiYurov2003,Morozov2024a,Morozov2024b,Morozov2026} for two-dimensional ideal hydrodynamics and related geophysical models \cite{Morozov2023b,Morozov2023c,Morozov2025,Morozov2026b}, as well as \cite{Morozov2026c} for the three-dimensional Euler--Helmholtz vorticity equation. 
When passing from inviscid Euler flow to viscous Navier--Stokes flow, we extend the  Lax system by introducing additional unknown fields whose compatibility conditions impose coupled equations on both the original and the newly introduced dependent variables. This approach is closely related to the methods of \emph{integrable deformation} \cite{Park1991} or \emph{integrable extension} \cite{Bogdanov2010,Bogdanov2017}, whose origins trace back to gauge-covariant Lax representations \cite{ZakharovShabat1979}. Related constructions have been extensively developed in the study of self-dual gravity, Einstein--Weyl geometry, and integrable equations on geometric backgrounds; see, in particular, \cite{Dunajski,BogdanovDryumaManakov} and the references therein, as well as work on the Manakov--Santini system \cite{ManakovSantini}.

In the present paper, we apply this extension method to construct Lax representations for the three-dimensional Navier--Stokes equations in vorticity form, starting from the Lax representation for the three-dimensional Euler--Helmholtz equation obtained in \cite{Morozov2026c}. Specifically, we enlarge the system of dependent variables by adjoining a new vector field $\vv$, which we call the \emph{vorticity-diffusion potential}.  The defining equation for $\vv$ is chosen so that the Lie bracket of $\crl \vv$ with the vorticity field $\ccrl \uu$ reproduces the viscous diffusion term $\Delta\,\ccrl \uu$. Geometrically, $\crl \vv$ acts as an auxiliary divergence-free ``diffusive drift velocity'': along the combined velocity field $\crl \uu + \crl \vv$, the vorticity is once again purely Lie-transported, just as in an inviscid flow. Furthermore,  the defining equation for $\vv$ is itself a curl-free condition. By the Poincar\'{e} lemma, it is therefore equivalent to the existence of an auxiliary scalar potential $p$, which is naturally incorporated into the Lax representation.

We also investigate the reduction of our three-dimensional construction to two spatial dimensions. Under translational symmetry along one coordinate axis, the vector potential $\uu$ reduces to the classical two-dimensional stream function $u$, and the three-dimensional Lax system reduces to a Lax representation for the two-dimensional Navier--Stokes vorticity equation. Moreover, we show that the reduced two-dimensional system admits a generalization involving an arbitrary smooth function $G$ of one variable, which can be viewed as a viscous integrable extension of the family of Lax representations for the two-dimensional Euler equations established in \cite{Morozov2024a}.


\section{Lax representation for the 3D Navier--Stokes equations}

\subsection{Preliminaries}

Let $M \subseteq \mathbb{R}^3$ be an open star-shaped domain with Cartesian coordinates $(x^1,x^2,x^3) = (x,y,z)$ and standard volume form $\mu = dx \wedge dy \wedge dz$. Let $\mathfrak{vect}(M)$  be the Lie algebra of smooth vector fields on $M$,
and let
\[
\mathfrak{svect}(M) = \{\pmb{a} \in \mathfrak{vect}(M) \,\,\vert\,\, L_{\pmb{a}} \mu = 0\}
= \{\pmb{a} \in \mathfrak{vect}(M) \,\,\vert\,\, \mathrm{div}\,\pmb{a} = 0\}
\]
denote the Lie subalgebra of volume-preserving (divergence-free) vector fields on $M$. The Lie algebra of outer derivations of 
$\mathfrak{svect}(M)$ is one-dimensional \cite{Lichnerowicz1974,Morimoto1976}; as its generator we take the Euler scaling (dilation) vector field
\begin{equation}
\ee = x\,\partial_x + y\,\partial_y + z\,\partial_z.
\label{Euler_field}
\end{equation}
Finally, to accommodate time-dependent fields, let $\mathfrak{g} = C^\infty(\mathbb{R}, \mathfrak{vect}(M))$ denote the current Lie algebra of smooth $\mathfrak{vect}(M)$-valued functions of time $t \in \mathbb{R}$, equipped with the pointwise Lie bracket.

\subsection{Vorticity formulation and the vorticity-diffusion potential}

Consider the motion of a viscous incompressible fluid in $M$. After normalizing the kinematic viscosity to unity by a suitable rescaling of variables, let $\ww \in \mathfrak{g}$ denote the fluid velocity field (up to sign convention) and $\crl \ww$ denote the vorticity field. Taking the curl of the momentum equation eliminates the hydrodynamic pressure and yields the 3D Navier--Stokes equations in vorticity form \cite{LandauLifshits}:
\begin{equation}
\crl \ww_t = [\ww, \crl \ww] + \Delta\,\crl \ww, 
\qquad \mathrm{div}\, \ww = 0,
\label{3D_NS_eq}
\end{equation}
where $\Delta \, a = a_{xx}+a_{yy}+a_{zz}$.

Since the domain $M$ is star-shaped and $\mathrm{div}\,\ww = 0$, the Poincar\'{e} lemma implies that there exists a vector potential $\uu \in \mathfrak{g}$ such that $\ww = \crl \uu$; see, e.g., \cite{Spivak_1}. Expressing the system in terms of $\uu$, the incompressibility constraint $\mathrm{div}\,\ww = \mathrm{div}(\crl \uu) = 0$ holds identically, the vorticity field becomes $\crl \ww = \ccrl \uu$, where we write $\ccrl \uu = \crl(\crl \uu)$, and the vorticity evolution equation in \eqref{3D_NS_eq} takes the form
\begin{equation}
\ccrl \uu_t = [\crl \uu, \ccrl \uu] + \Delta\,\ccrl \uu.
\label{main_eq}
\end{equation}

We introduce a new auxiliary vector field $\vv \in \mathfrak{g}$, which we call \emph{the vorticity-diffusion potential}, defined as a solution of the auxiliary equation
\begin{equation}
[\ccrl \uu, \crl \vv] = -\ccrl \Delta\, \uu.
\label{vorticity_diffusion_potential_def_eq}
\end{equation}
We can rewrite 
this equation as $\crl\big(\ccrl \uu \times \crl \vv - \Delta\,\crl \uu\big) = 0$.
By the Poincar\'{e} lemma on the star-shaped domain $M$, the vanishing of this curl is equivalent to the existence of a scalar potential $p$ whose gradient equals $\ccrl \uu \times \crl \vv - \Delta\,\crl \uu$.

\subsection{Lax representation in three dimensions}

The main result of this paper is formulated as follows:

\vskip 5 pt 
\noindent
{\sc Theorem 1}.  
{\it The extended Navier--Stokes system \eqref{main_eq}, \eqref{vorticity_diffusion_potential_def_eq} admits the Lax representation \begin{equation}
\left\{
\begin{array}{rcl}
\,\crl \qq_t &=& [\crl \uu +\crl \vv, \crl \qq] + [\ee,\crl \uu +\crl \vv],
\\
\,[\ccrl \uu, \crl \qq]&=& \lambda\,\ccrl \uu - [\ee,\ccrl \uu], 
\\
\,\mathrm{grad}\,p &=& \ccrl \uu \times \crl \vv -  \Delta\,\crl \uu,
\end{array}
\right.
\label{3D_Lax_representation}
\end{equation}
where $\qq \in \mathfrak{g}$ and $p \in C^\infty(\mathbb{R} \times M)$ are auxiliary dependent variables and 
$\lambda \in \mathbb{R}$ is a spectral parameter. That is, the compatibility conditions of the overdetermined system \eqref{3D_Lax_representation} with respect to $\qq$ and $p$ hold if and only if $\uu$ and $\vv$ satisfy system \eqref{main_eq}, \eqref{vorticity_diffusion_potential_def_eq}.
}

\vskip 5 pt
\noindent
{\it Proof.}
We verify the compatibility conditions of system \eqref{3D_Lax_representation} in two steps, corresponding to the elimination of the scalar potential $p$ and the vector potential $\qq$.

First, consider the third equation of \eqref{3D_Lax_representation}. On the star-shaped domain $M$, a necessary and sufficient condition for the existence of a smooth scalar function $p$ satisfying this relation is that the vector field on the right-hand side have vanishing curl:
\[
\crl\big(\ccrl \uu \times \crl \vv - \Delta\,\crl \uu\big) = 0.
\]
Since $\mathrm{div}(\ccrl \uu) = 0$ and $\mathrm{div}(\crl \vv) = 0$, 
we have $\crl(\ccrl \uu \times \crl \vv) = -[\ccrl \uu, \crl \vv]$. Using $\crl(\Delta\,\crl \uu) = \ccrl \Delta\,\uu$, we obtain
\[
-[\ccrl \uu, \crl \vv] - \ccrl \Delta\,\uu = 0,
\]
which is precisely the defining equation \eqref{vorticity_diffusion_potential_def_eq} for the vorticity-diffusion potential $\vv$.

Second, consider the first two equations of \eqref{3D_Lax_representation} for $\crl \qq$. To make their algebraic structure transparent, introduce the shifted vector field
\[
\hat{\qq} = \crl \qq - \ee.
\]
Since $\ee$ is independent of $t$ ($\ee_t = 0$) and $[\pmb{a}, -\ee] = [\ee, \pmb{a}]$ by skew-symmetry of the Lie bracket, the first two equations of \eqref{3D_Lax_representation} take the compact commutator form
\begin{equation}
\hat{\qq}_t = [\crl \uu + \crl \vv, \hat{\qq}],
\qquad
[\ccrl \uu, \hat{\qq}] = \lambda\,\ccrl \uu.
\label{shifted_Lax_3D}
\end{equation}
Differentiating the second equation of \eqref{shifted_Lax_3D} with respect to $t$ yields
\[
[\ccrl \uu_t, \hat{\qq}] + [\ccrl \uu, \hat{\qq}_t] = \lambda\,\ccrl \uu_t.
\]
Substituting $\hat{\qq}_t = [\crl \uu + \crl \vv, \hat{\qq}]$ from the first equation of \eqref{shifted_Lax_3D}, we have
\[
[\ccrl \uu_t, \hat{\qq}] + \big[\ccrl \uu, [\crl \uu + \crl \vv, \hat{\qq}]\big] = \lambda\,\ccrl \uu_t.
\]
By the Jacobi identity and the second equation of \eqref{shifted_Lax_3D},
\[
\big[\ccrl \uu, [\crl \uu + \crl \vv, \hat{\qq}]\big]
\]
\[
\qquad = -\big[[\crl \uu + \crl \vv, \ccrl \uu], \hat{\qq}\big]
+ \big[\crl \uu + \crl \vv, [\ccrl \uu, \hat{\qq}]\big]
\]
\[
\qquad = -\big[[\crl \uu + \crl \vv, \ccrl \uu], \hat{\qq}\big]
+ \lambda\,[\crl \uu + \crl \vv, \ccrl \uu].
\]
Now, using the vorticity-diffusion potential equation \eqref{vorticity_diffusion_potential_def_eq}  we set   
\[
\pmb{F} := \ccrl \uu_t - [\crl \uu + \crl \vv, \ccrl \uu]
\]
\[
\quad\, = \ccrl \uu_t - [\crl \uu, \ccrl \uu] - \Delta\,\ccrl \uu.
\]
Combining the preceding equations gives $[\pmb{F}, \hat{\qq}] = \lambda\,\pmb{F}$. Rewriting $\hat{\qq} = \crl \qq - \ee$, the compatibility condition becomes
\[
[\pmb{F}, \crl \qq] = \lambda\,\pmb{F} - [\ee, \pmb{F}].
\]
For this relation to hold identically for all solutions $\qq$ of the auxiliary system without imposing further differential constraints on $\qq$ and $\uu$, we must have $\pmb{F} = 0$, which is the 3D Navier--Stokes vorticity equation \eqref{main_eq}.
\hfill $\Box$


\section{Reduction to two spatial dimensions}

We now examine the symmetry reduction of the three-dimensional construction to flows that are independent of the vertical coordinate $z$. Specifically, we consider vector potentials and auxiliary fields of the form
\begin{equation}
\uu = - u\,\partial_z, \quad \vv = - v\,\partial_z, \quad \qq = - q\,\partial_z,
\quad u_z = v_z = q_z = p_z = 0,
\label{2D_ansatz}
\end{equation}
where $u, v, q, p$ are smooth scalar functions of $(t,x,y)$.

Substituting \eqref{2D_ansatz}  into \eqref{main_eq}, \eqref{vorticity_diffusion_potential_def_eq}, and \eqref{3D_Lax_representation} (and integrating the curl in the first equation of \eqref{3D_Lax_representation} by absorbing any additive function of $t$ into $q$), we obtain the two-dimensional Navier--Stokes vorticity equation
\begin{equation}
\Delta u_t = \J(u, \Delta u) + \Delta^2 u,
\label{2DNS}
\end{equation}
where $\J(a,b)=a_x b_y- a_y b_x$,
the two-dimensional vorticity-diffusion potential equation,
\begin{equation}
\J(\Delta u, v) = -\Delta^2 u;
\label{v_equation}
\end{equation}
and the reduced Lax representation for the coupled system \eqref{2DNS}, \eqref{v_equation}
\begin{equation}
\left\{
\begin{array}{rcl}
q_t &=& \J(u+v, q) + \E(u+v),
\\
\J(\Delta u, q) &=& \tilde{\lambda}\,\Delta u - \E(\Delta u), 
\\
p_x &=& - v_x \,\Delta u + \Delta u_y,
\\
p_y &=& - v_y \,\Delta u - \Delta u_x,
\end{array}
\right.
\label{2D_Lax_representation}
\end{equation}
where $\E(a) = x\,a_x +y\,a_y-2\,a$ and $\tilde{\lambda} = \lambda-1$.


\section{Generalization in two dimensions}

In two spatial dimensions, any smooth function $F(\Delta u)$ of the scalar vorticity satisfies $\J(\Delta u, F(\Delta u)) = 0$. This property allows us to generalize the reduced Lax representation \eqref{2D_Lax_representation} by replacing the linear term 
$\tilde{\lambda}\,\Delta u$ with an arbitrary smooth function of one variable. The resulting system provides a viscous integrable extension of the family of Lax representations for the two-dimensional Euler equations obtained in \cite{Morozov2024a}.

\vskip 5 pt
\noindent
{\sc Theorem 2.}
{\it The overdetermined system
\begin{equation}
\left\{
\begin{array}{rcl}
q_t &=& \J(u+v, q) + \varepsilon \,\E(u+v),
\\
\J(\Delta u, q) &=& \mu\,G(\mu^{-1}\,\Delta u + \lambda) - \varepsilon\, \E(\Delta u), 
\\
p_x &=& - v_x \,\Delta u + \Delta u_y,
\\
p_y &=& - v_y \,\Delta u - \Delta u_x,
\end{array}
\right.
\label{covering}
\end{equation}
where $\varepsilon \in \{0, 1\}$, $\lambda, \mu \in \mathbb{R}$ with $\mu \neq 0$, and $G$ is an arbitrary non-constant smooth function of one variable, defines a Lax representation for equations \eqref{2DNS} and \eqref{v_equation}.
}

\vskip 5 pt
\noindent
{\it Proof} \,\, is similar to the proof of Theorem 1.  
\hfill $\Box$

\section*{Acknowledgments}

I would like to express my sincere gratitude to I.S. Kra\-{}sil${}^{\prime}$\-{}shchik for very important discussions.


\bibliographystyle{amsplain}

\end{document}